\documentclass[prb,ams,amsmath,amssymb,twocolumn]{revtex4}
\usepackage{graphics}
\usepackage{dcolumn}
\usepackage{bm}
\input{epsf}
\newcommand{\beq}{\begin{equation}}
\newcommand{\eeq}{\end{equation}}
\def\la{\hbox{\raise.35ex\rlap{$<$}\lower.6ex\hbox{$\sim$}\ }}
\def\ga{\hbox{\raise.35ex\rlap{$>$}\lower.6ex\hbox{$\sim$}\ }}

\def\beq{\begin{equation}}
\def\eeq{\end{equation}}
\def\beqa{\begin{eqnarray}}
\def\eeqa{\end{eqnarray}}

\def\order#1{{\cal O}\left({#1}\right)}
\newcommand{\sfrac}[2]{ \mbox{$\frac{#1}{#2}$}}
\begin{document}

\baselineskip=12pt
\title{The instability of counter-propagating kernel gravity waves
in a constant shear flow}

\author{O.M. Umurhan}
\affiliation{Department of Geophysics and Planetary Sciences,
Tel--Aviv University, Israel}
\affiliation{Department of Physics, Technion-Israel Institute of
Technology, 32000 Haifa, Israel} \affiliation{Department of
Astronomy, City College of San Francisco, San Francisco, CA 94112,
USA  } \email[]{mumurhan@physics.technion.ac.il}

\author{E. Heifetz}
\affiliation{Department of Geophysics and Planetary Sciences,
Tel--Aviv University, Israel} \email[]{eyalh@cyclone.tau.ac.il}

\author{N. Harnik}
\affiliation{Department of Geophysics and Planetary Sciences,
Tel--Aviv University, Israel} \email[]{harnik@cyclone.tau.ac.il}

\author{F. Lott}
\affiliation{Laboratoire de Meteorologie Dynamique, Ecole Normale
Superieure, France}  \email[]{flott@lmd.ens.fr}

\date{\today}


\begin{abstract}
The mechanism describing the recently developed notion of kernel gravity waves (KGWs)
is reviewed and such structures are
employed to interpret the unstable dynamics of an example stratified plane parallel shear flow.
This flow has constant vertical shear, is infinite in the vertical extent, and characterized by
two density jumps of equal magnitude each decreasing successively with height,
in which the jumps are located symmetrically away
from the midplane of the system.  We find that for a suitably defined bulk-Richardson number
there exists a band of horizontal wavenumbers which exhibits normal-mode instability.
The instability mechanism closely parallels the mechanism responsible for the
instability seen in the problem of counter-propagating Rossby waves.
In this problem the instability arises out of the interaction of counter-propagating
gravity waves.  We argue that the instability meets the Hayashi-Young criterion
for wave instability.  We also argue that the instability is
the simplest one that can arise in a stratified atmosphere with constant shear flow.
The counter propagating gravity waves mechanism detailed here
explains why the Rayleigh criteria for shear flow instability in the
unstratified case does not need to be satisfied in the stratified case.
This illustrates how the Miles-Howard theorem may support destabilization
through stratification.
A normal mode analysis of a foamy layer consisting of two density jumps
of unequal magnitude is also analyzed.  The results are considered in terms
of observations made of sea-hurricane interfaces.
\end{abstract}

\maketitle
\section{Introduction}
Shear flows and the variety of instabilities they precipitate
are difficult processes to conceptualize despite over
a century of inquiry.  The two different interpretative tools
available are the theory of counter-propagating
Rossby Waves (CRWs)\cite{bretherton,heifetz99,heifetzCRW} and
classic over-reflection theory (O-R)\cite{lindzen} and these, in turn,
have been shown to be equivalent in rationalizing non-stratified shear
flows \cite{nili_n_eyal}.  While O-R theory can be used to interpret stratified shear flows,
the CRW approach cannot be used in the same way because
the basic building-block structures of CRW theory,
namely kernel Rossby waves (KRWs)\cite{heifetz99},
do not describe buoyancy.
However a formulation in the spirit of CRWs
has been recently developed
to include the effects of stratification\cite{HHUL07}.
The basic building blocks of this \emph{generalized wave kernel approach},
namely kernel Rossby-gravity waves (KRGWs), turn into
KRWs when buoyancy effects are absent.\par
We apply this approach to an example stratified shear flow in
order to observe how the mechanics of this interpretive tool
unfolds.  The geometry is an infinite atmosphere composed
of divergence free fluids in a globally constant shear flow.
The atmosphere is comprised of fluids of three densities
where the two density jumps of equal magnitude
are located symmetrically away from
the (nominal) midplane of the atmosphere.
\par
The important matter
to recognize is that because there is no-basic vorticity gradient
there will be no Rossby waves present at the density interfaces.
In this way the KRGW building blocks of the generalized theory will reduce
to what we call here \emph{kernel gravity waves} or ``KGWs" for short.
 {The KGW concept has antecedents tracing back to the work
of Sakai (1989)\cite{sakai} in which density-shear disturbances in a shallow-water
model are analyzed in terms
of \emph{physical wave coordinates}. The problem investigated by
Sakai \cite{sakai} cast the emerging instabilities in terms
of a wave interaction mechanism in which counter-propagating
waves influence each other from
a distance.  Baines and Mitsudera (1994)\cite{BM94} further
generalize these concepts as fitting into a more unified mechanism
of shear flow instabilities.  As an example they show how the
classic Holmboe instability \cite{holmboe62} may be rationalized in terms of
the interaction between the waves on separated
density and vorticity surfaces.}
\par
In the problem we examine here there will be four KGWs, two
associated at each interface. Of the two at each interface one
will travel faster than the local flow speed while the other will
be slower than the local flow speed. Of the four modes only two
modes, each counter-propagating with respect to the local flow
speed of its respective interface, may transit into a
stable-unstable pair. We find that for given values of the
bulk-Richardson number (defined below) there will always be a band
of horizontal wavenumbers which admit normal-mode instability. The
KGW analysis performed here shows something quite interesting: the
mechanics of the instability bears strong resemblance to the
mechanics responsible for the instability in the classic CRW
problem \cite{heifetz99} and, furthermore, become mathematically
equivalent in the limit of small horizontal wavelengths.  Thus
despite the fact that in this problem there are no Rossby waves
present (i.e. as edge waves) buoyancy oscillations in relative
motion with each other interact dynamically in such a way that it
can induce instability in the same way that CRWs produce
instability in Rayleigh's classic problem
\cite{heifetzCRW,Rayleigh1880}.\par Further reflection also shows
that onset of instability satisfies the criterion of Hayashi \&
Young (1987)\cite{hayashi_young}.  The stability criterion states
that for flows of this sort a transition into linear stability
occurs if there are individual waves in the flow which: (i)
propagate opposite to each other, (ii) have almost the same
Doppler-shifted frequency and (iii) can interact with one
another\cite{sakai}. Indeed in the current problem we see that
instability develops out of KGWs which counter-propagate at their
respective interface and through their action-at-a-distance
interaction \cite{heifetz99}  {trigger a transition into
instability for given values of a bulk
Richardson number.}
\par
We note that the work of Caulfield\cite{Caulfield94}, which
is an investigation of a variation of Holmboe's problem \cite{holmboe62}, contains as
a special limiting case the instability uncovered here.  In that
study a density configuration exactly like the one considered here
is analyzed.  The difference is that there exists symmetrically placed
jumps in the background vorticity resembling the classic Rayleigh profile.
These vorticity jumps, located at $z=\pm d$,
are at points some distance removed from the
density jumps located at $z=\pm h$.  In general many
types of modes appear in this problem.  Of these,
the ones referred to as Taylor modes may be shown
to limit to the type of modes
studied here when $d\gg h$.  However, this limiting
procedure was not performed and the qualitative significance
of this instability had not been appreciated at the time.
\par
It is important to note here that the three-layer
density configuration we consider in this work
has been proposed to be qualitatively applicable to
the question of drag reduction in rough seas
driven by hurricane force winds \cite{newell,powell}. The
middle ``foamy" layer \cite{Miki_2007} serves to dissipate
energy and transfer momentum between the air and sea.  Theoretical
work in problems with this sort of stratification
\cite{craik,Miki_2007} consider uniform
velocity profiles within each of the three layers but otherwise
different between them.  In this way we see the physical results
derived from this study as complementary to these cited works
because the velocity profile we consider has uniform shear
as opposed to (effective) delta-function jumps in the shear.
\par
This work is organized according to the following progr\'{a}mme.
In Section II we present and review the formulation derived
in Harnik et al. (2007)\cite{HHUL07}.  In Section III we work through the
theory for KGWs in an atmosphere of constant shear and a single density
jump.  As a matter of review we expend some effort
describing the mechanics of the KGW.  In Section IV we
introduce the problem of two density jumps of equal magnitude,
analyze its normal modes,
motivate its generalized formulation as an initial value problem and then argue for
the rationalizion of its
unstable dynamics.  We find that when the system is
unstable the mechanics of the modes
closely parallels the processes occurring for unstable
CRWs \cite{heifetzCRW} under conditions where the layers are not too close to one
another. We also spend a few words to rationalize the instability
in terms of the criterion of Hayashi \& Young (1987) and Sakai (1989)
\cite{hayashi_young,sakai}.
In Section V we study the normal-mode response of this
configuration when the layer is considered foamy. This
corresponds
to the problem considered in Section IV but where, instead, the density
jumps are not of equal magnitude. As in Shtemler et al. (2007) \cite{Miki_2007},
the density ratio of successive layers is measured by the parameter $\delta$ in
which $0<\delta<1$.  In comparison to the results of Section IV (where
$\delta \approx 1$),
we find that the range of unstable wavenumbers and the peak
growth rate shrinks as $\delta$ approaches zero.  We also find
that the modes are propagatory when they become unstable.
 {Section VI summarizes our results and we show that
the instability does not violate the Miles-Howard Theorem\cite{miles61,howard61}.
We also suggest that this instability, manifesting itself
under conditions that are classically considered to
be stably stratified, may be the simplest one possible
which can destabilize a plane-Couette profile.}
We conjecture
upon the results of this study and its relationship to the
problem of sea surface foam layers observed in hurricanes.

\section{Basic formulation: a review}

We begin the analysis by recasting the equations of linearized
motion in terms of
the formalism outlined in Harnik et al. (2007)\cite{HHUL07}.  Namely, the primitive equations of
motion describing Boussinessq incompressible 2D flow are
\beqa
\frac{Du}{Dt}  &=& - \bar U_z w-\frac{1}{\rho_0}\frac{\partial P}{\partial x}, \\
\frac{Dw}{Dt}  &=& -\frac{1}{\rho_0}\frac{\partial P}{\partial z} +b,
\eeqa
together with the equation of continuity and incompressibility
\beqa
\frac{Db}{Dt} &=& -w N^2,
\label{linearized_continuity_eqn} \\
\frac{\partial u}{\partial x} + \frac{\partial w}{\partial z} &=& 0.
\eeqa
The variables $u$ and $w$ are the horizontal ($x$) and vertical velocities ($z$)
 (respectively).  The horizontal mean flow is $\bar U$ and where we have defined
$\bar U_z \equiv \frac{dU}{dz}$ as the mean shear together
with
$\frac{D}{Dt} \equiv (\partial_t + \bar U\partial_x)$.
The density $\rho$ is cast in terms of the variable $b\equiv - g\rho/\rho_0$
and where
\[N^2 \equiv   -\frac{g}{\rho_0}\frac{d\rho}{dz} = \frac{d\bar b}{dz}. \]
Because the flow is incompressible
the evolution of the density $\rho(z,t)$ may be equivalently traced by the motion
of some material invariant.
In this case we follow the vertical displacement $\zeta$ of
a fluid parcel which was otherwise at rest.  The undisturbed density profile is assumed to have
a vertical variation of some sort and, thus, the displacement may be easily
associated with clearly identifiable surfaces.
Thus, as was demonstrated in Harnik et al. (2007)\cite{HHUL07}, we shall write the equations of motion
in terms of the evolution of the vorticity and vertical displacement as
\beqa
\frac{Dq}{Dt} &=& - \bar q_z w + \frac{\partial b}{\partial x} , \label{q_eqn}\\
\frac{D\zeta}{Dt} &=& w(x,z), \label{zeta_eqn}
\eeqa
where $\bar q_z \equiv - \bar U_{zz}$, $b = -\bar b_z \zeta$ and where,
\beq
w(x,z) = \frac{\partial\psi}{\partial x},
\qquad
u(x,z) = -\frac{\partial\psi}{\partial z},
\label{streamfunction_solution}
\eeq
in which we have expressed velocities in terms of the usual
stream function formulation of 2D incompressible problems
and where the vorticity $q$ is
\beq
q = \frac{\partial w}{\partial x}  - \frac{\partial u}{\partial z} = \nabla^2 \psi
\label{streamfunction_eqn}
\eeq
\section{Single density jump}
From here on out we assume the background flow state to be a
constant shear (i.e. $\bar q_z = \bar U_{zz} = 0$). We examine here a single density interface located
at the position $z=0$. The basic state density is written as
\beq \bar\rho =
\rho_0 + {\Delta\bar\rho} H(z), \eeq
where $H$ is the Heaviside
function.
The equations of motion now become,
\beqa
(\partial_t + U\partial_x) q &=& (g\Delta\rho/\rho_0)\partial_x
\zeta \delta(z) \label{vorty_eqn}
, \\
(\partial_t + U\partial_x)\zeta &=& w(x,0).
\eeqa
(\ref{vorty_eqn})
says that the vorticity ought to also have a delta function dependence.
Thus we make the ansatz
\[
q = \hat q(x,t)\delta(z).
\]
To proceed, we assume that all variables have a Fourier
decomposition in the $x$ direction, that is to say
\[
\sim e^{ikx} + {\rm c.c.}
\]
where $k$ is the horizontal wavenumber.
We begin with the solution to the streamfunction.  As the ansatz
indicates, away from the interfaces the vorticity is zero.  Thus
it means that one formally writes the solution to the stream function
equation (\ref{streamfunction_eqn})
as
\beq
\psi = \hat q(x,t) \int_{-\infty}^{\infty}G(z,z')dz',
\eeq
since $\nabla^2 G = (\partial_z^2 - k^2) G(z,z') = \delta(z-z_0)$,
and because this geometry is so simple the solution for $\psi$
is
\beq
\psi = -\frac{\hat q}{2k} e^{-k|z-z_0|}.
\eeq
Implicit in the construction of this solution is that (a) the vertical velocity,
$w = \partial_x \psi$ is
continuous across the interface and (b)
the vorticity delta function results from the jump in u across $z_0$.
We are now in a position to put these expression explictly into the equations of
motion above.  Thus one has, after shaking out algebraic factors and the solution ansatz,
\beqa
(\partial_t + ik U_0) \hat q &=& ik\frac{g\Delta\rho}{\rho_0}\zeta, \\
(\partial_t + ik U_0)\zeta &=&-\frac{i}{2}\hat q,
\eeqa
where $U_0 = U(z=0)$.  Let us suppose that $U_0 = 0$ for this problem.
Let us define the gravity wave speed $c_g \equiv N/k = \sqrt{-\frac{g\Delta\bar\rho}{2\rho_0 k}}$.
Provided that $\Delta\bar\rho <0$ (stably stratified) then the gravity wave speed
is real - and this we assume henceforth.
We find that the above two equations appear rewritten as
\beqa
\partial_t\hat q &=& -2ik^2c_g^2\zeta, \\
\partial_t\zeta &=&-\frac{i}{2}\hat q.
\eeqa
The above is the ode system, in time, describing the evolution of
simple classical Rayleigh-Taylor modes. Note that this system is
normal as far as the operators are concerned \cite{Schmid}.
We can further proceed by defining new quantities
through the relationship
\beq
\zeta_{\pm} =
\frac{1}{2}\left(\zeta \pm \frac{1}{2kc_{g}}\hat q\right),
\eeq
revealing to us the simplified system
\beqa
(\partial_t + ik c_{g})\zeta_+ &=& 0, \\
(\partial_t - ik c_{g})\zeta_- &=& 0.
\eeqa
A normal mode analsysis may proceed from here by assuming $e^{-ikct}$ dependence
upon all the solutions in which $c$ is the wavespeed.
It is quite straightforward to establish that the eigenmodes for this system are
described by
\beqa
(c-c_g)\zeta_+ &=& 0 \label{mode1}, \\
(c+c_g)\zeta_- &=& 0\label{mode2}.
\eeqa
For the sake of this illustration let us consider the
propagation of the $\zeta_+$ mode
which propagates at speed $c=c_g$.  To study
$\zeta_+$ in isolation means setting
$\zeta_-$ to zero which, in turn, means that,
\[
\zeta \sim \hat q.
\]
In other words, the disturbance height is positively correlated
with the local vorticity.\par
\begin{figure}
\begin{center}
\leavevmode
\epsfysize=6.5cm
\epsfbox{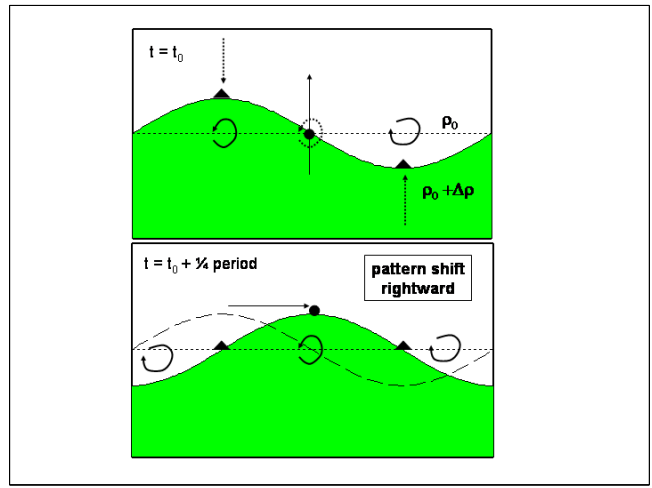}
\end{center}
\caption{{\small
A mechanism schematic for the single kernel gravity wave $\hat\zeta_+$
in which positive disturbances are correlated with positive vorticity
disturbances.
Fluid of two differing densities are disturbed with sinusoidal profile.
The corresponding negative/positive vorticity profile is represented by solid clockwise/counterclockwise
arrows.  Restricting attention to between the two labeled
triangles in the top panel, the
density arrangement looks like a displaced lever arm in a gravitational
field whose 1D pivot (fulcrum) point
is located
at the position of the solid circle.  This physically
represents the tendency for the fluid to flatten out.
The lever arm effect has
a tendency to create positive vorticity about
the fulcrum (dashed counterclockwise arrow).
Concurrently, the
preexisting flow due to the disturbance profile (solid
clockwise/counterclockwise arrows) causes this
point to rise upwards (solid arrow).  As vorticity is created there, it also
begins
to create
a velocity field at the triangles causing those positions
to move (according to the manner denoted by the dashed arrows).
By the time the fulcrum point reaches its peak, due to the lever
arm effect there now exists positive vorticity in this local
patch of the fluid after
one quarter period.  The pattern consequently shifts rightward
(bottom panel).
}}
\label{mechanism_plots}
\end{figure}

Referring to Fig. \ref{mechanism_plots} we  depict
this situation for a stably stratified density profile.
We focus our attention to the section of fluid located
between the two triangles (representing the peak and trough
of the wave).  The fluid layer has been disturbed from
its equilibrium configuration (denoted by dashed lines) in such
a way that it looks like a lever arm out of balance with the
fulcrum denoted with a solid filled circle.  One's intuition
instructs that in this condition the lever
arm has the tendency to rotate in a counter
clockwise sense (positive vorticity) about this fulcrum point.
 {This is simply the fluid's tendency to flatten out.}
Thus the out-of-equilibrium arrangement equates to a local
creation of vorticity, i.e. a positive time rate of change of
vorticity there (shown with a dashed counter-clockwise arrow on
the graph).  However, the already preexisting vorticity of this
disturbance (shown with solid counter-clockwise arrow) causes, for
example, the position of the fulcrum point to move up because the
velocity field there points upwards (solid arrow).  As this
section of fluid responds, vertical velocity fields also develop
at the positions of the triangles in proportion to the amount of
vorticity being generated at the fulcrum (denoted with the dashed
vertical arrows above and below the left and right triangles
respectively).\par In the panel immediately below we see what
happens after one quarter cycle: the positive time rate of change
of vorticity at the point which was once the fulcrum has turned
into a positive vorticity and the fulcrum point has turned into
the maximum height of the wave.  As this occurred, the position of
the triangles have moved correspondingly.  In this way the entire
pattern has shifted to the right.   {The mechanics of the
oppositely propagating mode, $\zeta_-$, are recovered in a simple
way: $q$ and correspondingly $w$ are reversed and hence
$\partial_t q$ and $\partial_t\zeta$ are then also reversed and
then, consequently, the pattern moves to the left.  Note that in
this case $\zeta_-$ and $q$ are in phase.}

\section{Two density jumps in a globally constant shear}
We now consider a situation in which there are
two density jumps located symmetrically
at a distance $h$ away from the position $z=0$.  Thus,
\beq
\bar\rho = \Delta\bar\rho_1 H(z-z_1) + \rho_0 + \Delta\bar\rho_2H(z-z_2),
\label{density_profile}
\eeq
where the locations $z_{1,2}$ are given by
\[
z_1 = h + \zeta_1(x,t), \qquad
z_2 = -h + \zeta_2(x,t).
\]
We assume a configuration in which the shear is globally constant so that
$\bar U = \Lambda z$ implying that $\bar q_z = 0$.
We further say that $\Lambda \equiv  U/h$ where
$U$ is the magnitude of the background flow at $z=\pm h$.
We have written
matters in this way to always make sure that at $z=h$ the wind is
easterly (positive), in other words, we will assume that $U
>0$ without loss of generality.  Also, as before, we shall consider
stably stratified configurations so that both
$\Delta\bar\rho_1 < 0$ and
$\Delta\bar\rho_2 < 1$.
\par
Inspection shows that this
form of the unperturbed density spawns delta function forms for the
vorticity as we encountered before.  In other words it means that we write
\beq
q =  \hat q_1(x,t) \delta(z-z_1) +  \hat q_2(x,t)\delta(z-z_2).
\eeq
Putting these forms into the governing equations (\ref{q_eqn}-\ref{zeta_eqn}) and
assuming the single-wave Fourier ansatz
\beq
\{\hat q_1,\zeta_1,\hat q_2,\zeta_2,\psi\}^{\bf T} =
\{\tilde q_1,\tilde \zeta_1,\tilde q_2,\tilde\zeta_2,
\tilde\psi\}^{\bf T} e^{ikx} + {\rm c.c.},
\label{fourier_ansatz}
\eeq
we find the following four equations to emerge,
\beqa
\left(\frac{\partial}{\partial t} + ik U\right)\tilde q_1 &=& ikh \tilde N^2_{_1}\tilde\zeta_1, \label{hat_q1_eqn}\\
\left(\frac{\partial}{\partial t} + ik U\right)\tilde\zeta_1 &=& \tilde w(h), \\
\left(\frac{\partial}{\partial t} - ik U\right)\tilde q_2 &=& ikh \tilde N^2_{_2}\tilde\zeta_2, \\
\left(\frac{\partial}{\partial t} - ik U\right)\tilde\zeta_2 &=& \tilde w(-h).
\label{hat_zeta2_eqn}
\eeqa
with $\tilde N^2_{_{1,2}} \equiv-\frac{g\Delta\rho_{1,2}}{h\rho_0}$.  We refer the
reader to Appendix A for a detailed derivation of these equations from first principles.
The solution to the streamfunction, i.e. (\ref{hat_psi_solution}), is
rewritten here:
\beq
\tilde\psi = -\frac{\tilde q_1}{2k}e^{-k|z-h|}
-\frac{\tilde q_2}{2k}e^{-k|z+h|}.\eeq
  Since $\tilde w \rightarrow ik\tilde\psi$ we may explicitly
  write the vertical velocity at the levels $z=\pm h$
\[
\tilde w(h) =
-\frac{i\tilde q_2}{2}e^{-2kh}
-\frac{i\tilde q_1}{2}, \qquad
\tilde w(-h) = -\frac{i\tilde q_2}{2}
-\frac{i\tilde q_1}{2}e^{-2kh}.
\]
Thus, we have finally
\beqa
\left(\frac{\partial}{\partial t} + ik U\right)\tilde q_1 &=& -2ik^2c_{g1}^2\tilde\zeta_1, \\
\left(\frac{\partial}{\partial t} + ik U\right)\tilde\zeta_1 &=& -i\frac{\tilde q_2}{2}e^{-2kh}
-i\frac{\tilde q_1}{2}, \\
\left(\frac{\partial}{\partial t} - ik U\right)\tilde q_2 &=& -2ik^2c_{g2}^2\tilde\zeta_2, \\
\left(\frac{\partial}{\partial t} - ik U\right)\tilde\zeta_2 &=& -i\frac{\tilde q_2}{2}
-i\frac{\tilde q_1}{2}e^{-2kh},
\eeqa
in which
\[
c_{g1}^2 \equiv -\frac{g\Delta\rho_1}{2\rho_0k}, \qquad
c_{g2}^2 \equiv -\frac{g\Delta\rho_2}{2\rho_0k}.
\]

\par
 We
may recast the equations of motion in this case in terms of the individual
``kernel" waves that propagate on each interface (see Section III and
Harnik et al. 2007\cite{HHUL07}).  This re-formulation
will be beneficial because it will show us transparently how these individual
waves interact with one another.  To that end we define the correlated/anticorrelated (``$\pm$")
wave quantities appropriate to each surface,
\beqa
\tilde\zeta_{_{1\pm}} &=& \frac{1}{2}\left(\tilde\zeta_{1} \pm \frac{1}{2kc_{g1}}\tilde q_1\right), \\
\tilde\zeta_{_{2\pm}} &=& \frac{1}{2}\left(\tilde\zeta_{2} \pm \frac{1}{2kc_{g2}}\tilde q_2\right),
\eeqa
substituting these definitions into (\ref{normal_mode_equations})
along with the observation that
\[
\frac{\tilde q_1}{2kc_{_{g1}}} = \tilde\zeta_{_{1+}}-\tilde\zeta_{_{1-}},
\qquad
\frac{\tilde q_2}{2kc_{_{g2}}} = \tilde\zeta_{_{2+}}-\tilde\zeta_{_{2-}},
\]
and after some reshuffling of terms this results in,
\begin{subequations}
\label{equations_recast}
\beqa
\left(\frac{\partial}{\partial t}+ikc_{{g1}}+ikU\right)\tilde\zeta_{_{1+}} &=& -ik\beta_2
\Bigl(\tilde\zeta_{_{2+}}-\tilde\zeta_{_{2-}}\Bigr), \ \ \\
\left(\frac{\partial}{\partial t}-ikc_{{g1}}+ikU \right)\tilde\zeta_{_{1-}} &=& -ik\beta_2
\Bigl(\tilde\zeta_{_{2+}}-\tilde\zeta_{_{2-}}\Bigr), \ \ \label{counterKGW_top_eqn}
\\
\left(\frac{\partial}{\partial t}+ikc_{{g2}}-ikU \right)\tilde\zeta_{_{2+}} &=& -ik\beta_1
\Bigl(\tilde\zeta_{_{1+}}-\tilde\zeta_{_{1-}}\Bigr), \ \
\label{counterKGW_bottom_eqn}
\\
\left(\frac{\partial}{\partial t}-ikc_{{g2}}-ikU\right)\tilde\zeta_{_{2-}} &=& -ik\beta_1
\Bigl(\tilde\zeta_{_{1+}}-\tilde\zeta_{_{1-}}\Bigr), \ \
\eeqa
\end{subequations}
where
\[
\beta_1 = c_{_{g1}}\frac{e^{-2kh}}{2};\qquad
\beta_2 = c_{_{g2}}\frac{e^{-2kh}}{2}.
\]
Provided that $c_{{g1}}$ and $c_{{g2}}$ are real (that is to say, that the
density succesively gets smaller with increased height)
, the structures $\tilde\zeta_{_{1\pm}},\tilde\zeta_{_{2\pm}}$ are each interpreted as pairs
of kernel gravity
wave modes at their respective surfaces.  Note that $\tilde\zeta_{_{1+}}$
and $\tilde\zeta_{_{2-}}$ are waves which move in the same direction as the local
flow speed (at their respective levels) while
$\tilde\zeta_{_{1-}}$
and $\tilde\zeta_{_{2+}}$ move against the local flow speed.  The latter two structures
are central actors when instability develops (see below).

\subsection{Normal Modes - and Recasting}
Now we can take this further and assume $e^{-ikct}$ solutions and inquire
about the normal mode behavior of (\ref{equations_recast}), revealing
\begin{subequations}
\label{normal_mode_equations}
\beqa
(c-c_{{g1}}-U )\tilde\zeta_{_{1+}} &=& \beta_2
\Bigl(\tilde\zeta_{_{2+}}-\tilde\zeta_{_{2-}}\Bigr),\\
(c+c_{{g1}}-U )\tilde\zeta_{_{1-}} &=& \beta_2
\Bigl(\tilde\zeta_{_{2+}}-\tilde\zeta_{_{2-}}\Bigr),\\
(c-c_{{g2}}+U )\tilde\zeta_{_{2+}} &=& \beta_1
\Bigl(\tilde\zeta_{_{1+}}-\tilde\zeta_{_{1-}}\Bigr),\\
(c+c_{{g2}}+U )\tilde\zeta_{_{2-}} &=& \beta_1
\Bigl(\tilde\zeta_{_{1+}}-\tilde\zeta_{_{1-}}\Bigr).
\eeqa
\end{subequations}



\begin{figure}
\begin{center}
\leavevmode
\epsfysize=6.5cm
\epsfbox{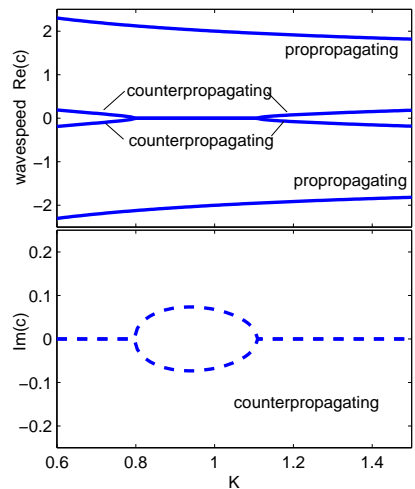}
\end{center}
\caption{{\small
The dispersion relationship (\ref{dispersion_relation}) for the case where $c_{g1}=c_{g2}$
with $Ri=1$, i.e. (\ref{nice_dispersion_relationship}).  The top graph shows the real wavespeeds of the
four
modes as a function of $K$.  The propropagating modes always remain
stable while
the two counter-propagating modes become unstable when they phase lock
with each other and their phase speeds become zero.  The bottom graph
shows the imaginary wavespeeds of the counter-propagating modes.
}}
\label{total_dispersion}
\end{figure}

\begin{figure}
\begin{center}
\leavevmode
\epsfysize=6.5cm
\epsfbox{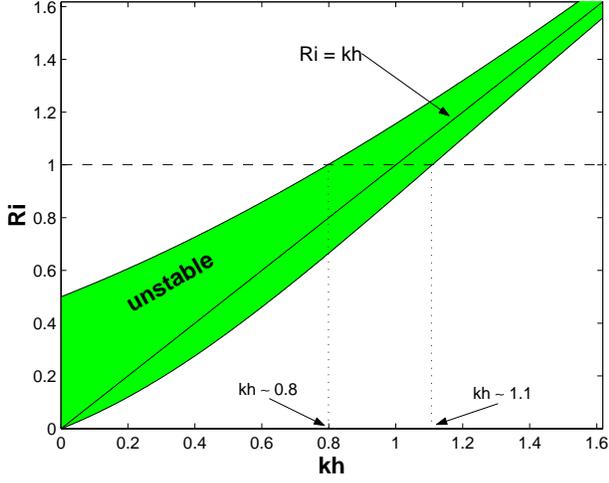}
\end{center}
\caption{{\small
The incidence of
linear instability (shaded region) for given values of the
bulk-Richardson Number Ri and the wavenumber $k$.  In this particular
example $c_{_{g1}} = c_{_{g2}}$.  Shown with a dashed horizontal
line is the specific value of Ri $= 1$ that is examined in the text.  Instability occurs
 approximately for the range $0.80 < kh < 1.11$ (shown with dotted lines).
The range of instability shrinks as $kh \gg 1$ and stradles
the line Ri $=kh$ (also depicted).
}}
\label{linear_instability_diagram}
\end{figure}

One can quite easily generate a dispersion relation for the above
set
\beqa
& & c^4 - (c_{g1}^2 + c_{g2}^2 + 2U^2)c^2 + 2U(c_{g2}^2-c_{g1}^2)c + \nonumber \\
& & \ \ \  U^4 - U^2(c_{g1}^2+c_{g2}^2) + c_{g1}^2c_{g2}^2\left(1-e^{-4kh}\right) = 0. \ \ \ \ \
\label{dispersion_relation}
\eeqa
From here on out we shall consider the simple profile
$c_{g1}^2 = c_{g2}^2 = c_g^2.$, i.e. that
the drop in the density profile is uniform from one step
to the next, i.e. that $\Delta\bar\rho_{1} = \Delta\bar\rho_{2} = -\Delta\rho$ where
$\Delta\rho >0$.  We write the gravity wave speed as $c_g^2$ which
is now assumed to be greater than zero (i.e. a real wavespeed).
The solution to the wavespeeds becomes
\beqa
\left(\frac{c_{_\pm}}{\bar U}\right)^2 &=& (1 \pm {\cal C})^2 \pm {\cal C}_{\rm ind}^2; \nonumber \\
   {\cal C}_{\rm ind}^2 &\equiv&  2{\cal C}
\left ( \sqrt{1+\frac{{\cal C}^2e^{-4K}}{4}} - 1 \right),
\label{nice_dispersion_relationship}
\eeqa
where $K\equiv kh$ and
we have defined the scaled gravity wave speed with respect to the
local shear velocity as
\beq
{\cal C}^2 = \frac{c_g^2}{U^2} = \frac{Ri}{K}, \qquad
Ri \equiv \left(\frac{gh\Delta\rho}{2\rho_0 U^2}\right),
\eeq
where $Ri$ is interpreted as the (bulk) Richardson number.  We defined an ``induced"
wavespeed ${\cal C}_{\rm ind}$ and its significance will be explained
shortly.
In general this system has four modes given by $\pm c_{_\pm}$
which are depicted in Figure \ref{total_dispersion} for bulk Richardson number $Ri = 1$.
Inspection of the above shows that two roots are always
real (the ``+" branch) and we identify these as being
propropagating modes while the negative branch
can admit an exponentially decaying/growing pair and
we refer to these as counterpropagating modes.
This exists for a band of gravitywave speeds
i.e. that is when the condition
\[
1 + {\cal C}^2 - 2{\cal C} \sqrt{1+\frac{{\cal C}^2e^{-4K}}{4}} < 0,
\]
is satisfied - and is so when
\beq
\frac{K}{1-e^{-2K}} > Ri >\frac{K}{1+e^{-2K}}.
\label{Ri_K_relationship}
\eeq
Figure \ref{linear_instability_diagram} diagramatically encapsulates
the result in expression (\ref{Ri_K_relationship}).
There are two things to note.  First is that for any given Ri there always
exists a band of wavenumbers supporting an instability.  Secondly, inspection
of (\ref{Ri_K_relationship}) quickly shows that this band centers around
the line Ri $=K$ and that, further, the size of this band becomes
exponentially small as $K\gg 1$.  In other words, as $K\rightarrow \infty$
the range of instability occurs within a small region centered
on $K$ whose width is measured by $\delta \equiv
 e^{-2K}$, i.e.
\[
{K}(1-\delta) < Ri < {K}(1+\delta).
\]
\par
\begin{figure}
\begin{center}
\leavevmode
\epsfysize=9.cm
\epsfbox{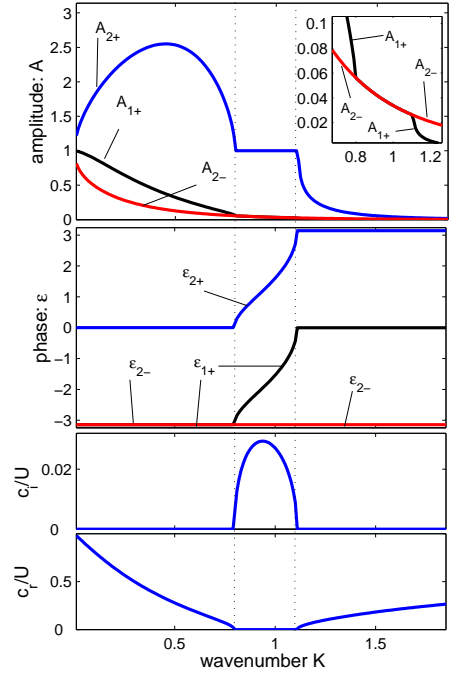}
\end{center}
\caption{{\small Phase and amplitude diagram for the unstable
mode ($+c_{_{-}}$) in which $Ri = 1$. All kernel wave phases and amplitudes
are measured against the wave $\tilde\zeta_{_{1-}}$ which is set with
$A_{_{1-}}=1, \epsilon_{_{1-}}=0$.
The panels on the right shows
the complex wave speeds for the unstable mode ($+c_{_{-}}$)
as a function of $K$.
The band of instability
in $K$ is delineated with dotted lines.  For $K\gg 1$ all
mode amplitudes go to zero leaving behind only the wave $\tilde\zeta_{_{1-}}$.
In the unstable regime $\zeta_{_{2+}}$ shares the same amplitude
as $\tilde\zeta_{_{1-}}$
but begins to show an eastward tilt (in the same
sense of the shear) of its phase, $\epsilon_{_{2+}} >0$
as one sweeps thru $K$.  In the unstable phase, the amplitudes
of the other two modes (i.e. $\tilde\zeta_{_{1+}},\tilde\zeta_{_{2-}}$)
also lock together and there is also
an eastward tilt
 of their phases with respect to each other in this
range (i.e. of $\epsilon_{_{1+}}$ with respect to $\epsilon_{_{2-}}$).  Note
that the amplitudes of $A_{_2-}$ and $A_{_{1+}}$
in this regime are small compared to the other two amplitudes.
The circles represent the function $\beta\sin(\epsilon_{_{2+}}-\epsilon_{_{1-}})$.
}}
\label{phase_diagram_unstable}
\end{figure}

We ask the question: for the
unstable mode in isolation, what are the
relative amplitudes and phases of the four
kernel waves with respect to each other?
Let us say define the vector $\mathbf{V} = (\tilde\zeta_{_{1+}},\tilde\zeta_{_{1-}},
\tilde\zeta_{_{2+}},
\tilde\zeta_{_{2-}})^\mathrm{T}$, then we see
that (\ref{normal_mode_equations}) can be rewritten as
\beq
c\mathbf{V} = \mathbb{L}\mathbf{V}
\label{linear_theory_compact_form}
\eeq
in which $\beta \equiv {c_g}e^{-2K}/2$ and
\beq
\mathbb{L} \equiv \left(
\begin{array}{cccc}
c_{g}+U& 0 & \beta & -\beta \\
0& -c_{g}+U & \beta & -\beta  \\
 \beta & -\beta  & c_{g}-U& 0 \\
 \beta & -\beta  & 0 &-c_{g}-U
\end{array}
\right).
\eeq
In the diagonal basis (denoted by primes) this system may be rewritten as
\beq
c\mathbf{V}' = \mathbb{C}\mathbf{V}', \qquad
\mathbb{C} \equiv diag(+c_{_-},-c_{_-},+c_{_+},-c_{_+}).
\eeq
where $\mathbf{V}' = \mathbb{R}\mathbf{V}$ for the rotation
vector $\mathbb{R}$.  The columns of the inverse
rotation matrix $\mathbb{R}^{-1}$ are composed of
the eigenvectors of $\mathbb{L}$ (respecting the order of the eigenvalues
appearing in
$\mathbb{C}$).  We have written the matrix $\mathbb{C}$ so that the
first position represents the mode that can go unstable.  Thus, if
we wish to observe this mode in isolation we must require
the following three equations
\beq
\mathbb{R}_{2n}\mathbf{V}_n = 0,\qquad
\mathbb{R}_{3n}\mathbf{V}_n = 0,\qquad
\mathbb{R}_{4n}\mathbf{V}_n = 0,
\label{unstable_mode_behavior}
\eeq
($n = 1,2,3,4$) be simultaneously satisfied.  (Note that
this is a generalization of the wave-mode isolation procedure
we performed in Section III to follow the single
kernel wave $\tilde\zeta_{+}$.)
Since these
are three equations with four unknowns, we ask what
are the relative amplitudes and phases of the
remaining three kernel waves when one has been
set fixed to $1$.  In this particular case of the unstable
mode, we find that the relative behavior is most
transparently appreciated when we gauge the action of
the kernel waves with respect to the kernel wave
$\tilde\zeta_{_{1-}}$ - i.e. the mode which, in the limit
of large separation, corresponds to the
wave propagating against the mean flow at the upper
interface, cf. (\ref{normal_mode_equations}c) with $kh \gg 1$.
By choosing $\tilde\zeta_{_{1-}} = 1$ we can solve
(\ref{unstable_mode_behavior}) for the remaining
waves.  Although $\mathbb{R}^{-1}$ may be
written down analytically the solution for $\mathbb{R}$
is quite cumbersome and, as such, we depict a typical example
numerically
of the resulting behavior in Fig. \ref{phase_diagram_unstable}.
To be specific, we write each kernel wave in terms
of its amplitude and phase, i.e.
\[
\tilde\zeta_{_{1\pm}} = A_{_{1\pm}}e^{i\epsilon_{_{1\pm}}},\qquad
\tilde\zeta_{_{2\pm}} = A_{_{2\pm}}e^{i\epsilon_{_{2\pm}}}.
\]
By construction $A_{_{1-}}=1, \ \epsilon_{_{1-}}=0$.  In the figure
we plot the relative amplitudes and phases for the remaining waves
with respect to $\tilde\zeta_{_{1-}}$ as one sweeps through wavenumber $K$.
\par
We first note that in the limit of large
wavenumber $\beta$ becomes negligible
and the four equations of
(\ref{linear_theory_compact_form}) decouple.  This means
to say that the four kernel modes propagate not ``knowing"
of the other waves or interfaces.
This trend
is represented in Fig.\ref{phase_diagram_unstable} for $kh \gg 1$ -
which shows the amplitudes of the remaining
modes (i.e. $\tilde\zeta_{_{1+}},\tilde\zeta_{_{2\pm}}$) to go
to zero as well.
In the unstable zone (roughly $0.8<K<1.2$) the amplitudes
of $\tilde\zeta_{_{1-}}$ and $\tilde\zeta_{_{2+}}$ stay equal, i.e.
$A_{_{1-}}=A_{_{2+}} = 1$, and the same goes for the
the other two modes $A_{_{1+}}=A_{_{2-}}$ although they
do not remain constant in the instability range as do the other pair.
For this particular set of parameters the amplitudes of
$A_{_{1+}},A_{_{2-}}$ are quite small compared to $A_{_{1-}}$
and this feature can aid us in developing a mechanical rationalization of
the instability (see next section).  We note that the phases
also show an eastward tilt with respect to each other in the unstable
regime:  $\epsilon_{_{2+}}$ begins to show a positive phase with respect
to $\epsilon_{_{1-}} = 0$ as well as $\epsilon_{_{1+}}$ showing a positive
phase with respect to $\epsilon_{_{2-}} = -\pi$ (the latter, incidentally, stays
fixed throughout).\par
Due to the symmetries of the normal modes
we note that the action and relative configuration of the kernel waves for the stable mode
(i.e. $-c_{_-}$) is also represented in the figure with two modifications:
the first of these is to set $c_{_-} \rightarrow -c_{_{-}}$ while
the second would be to interchange the roles
between $\tilde\zeta_{_{1-}}$ and $\tilde\zeta_{_{2+}}$
and between
$\tilde\zeta_{_{2-}}$ and $\tilde\zeta_{_{1+}}$.
In the exponentially decaying regime
$\tilde\zeta_{_{2+}}$ starts to show a westward drift of its phase with respect
to $\tilde\zeta_{_{1-}}$.

\subsection{Formulation as an initial value problem}
\bigskip
The equations describing the general linearized response also offers insights
which will aid us in rationalizing the instability uncovered here.
Thus we
consider the general temporal response of the equations of motion (\ref{equations_recast})
by adopting the strategy of Heifetz \& Methven (2005)\cite{heifetzCRW} and inserting
into those equations the assumed forms
\beqa
& & \tilde\zeta_{_{1+}} = Z_{_{1+}}(t)e^{i\varepsilon_{_{1+}}(t)},\quad
\tilde\zeta_{_{1-}} = Z_{_{1-}}(t)e^{i\varepsilon_{_{1-}}(t)}, \nonumber \\
& & \tilde\zeta_{_{2+}} = Z_{_{2+}}(t)e^{i\varepsilon_{_{2+}}(t)},\quad
\tilde\zeta_{_{2-}} = Z_{_{2-}}(t)e^{i\varepsilon_{_{2-}}(t)}.
\eeqa
After equating real and imaginary parts we have the eight ode's: four for the amplitudes,
\begin{subequations}
\label{eight_odes}
\beqa
\dot Z_{_{1+}} &=& k\beta\left[Z_{_{2+}}\sin\bigl(\varepsilon_{_{2+}}-\varepsilon_{_{1+}}\bigr)
-
Z_{_{2-}}\sin\bigl(\varepsilon_{_{2-}}-\varepsilon_{_{1+}}\bigr)
\right]
,\\
\dot Z_{_{1-}} &=& k\beta\left[Z_{_{2+}}\sin\bigl(\varepsilon_{_{2+}}-\varepsilon_{_{1-}}\bigr)
-
Z_{_{2-}}\sin\bigl(\varepsilon_{_{2-}}-\varepsilon_{_{1-}}\bigr)
\right]
,\\
\dot Z_{_{2+}} &=& k\beta\left[-Z_{_{1+}}\sin\bigl(\varepsilon_{_{2+}}-\varepsilon_{_{1+}}\bigr)
+
Z_{_{1-}}\sin\bigl(\varepsilon_{_{2+}}-\varepsilon_{_{1-}}\bigr)
\right]
,\\
\dot Z_{_{2-}} &=& k\beta\left[-Z_{_{1+}}\sin\bigl(\varepsilon_{_{2-}}-\varepsilon_{_{1+}}\bigr)
+
Z_{_{1-}}\sin\bigl(\varepsilon_{_{2-}}-\varepsilon_{_{1-}}\bigr)
\right]
,
\eeqa
and four for the respective phases,
\beqa
& & \dot \varepsilon_{_{1+}} = -k(U+c_g) \nonumber \\
& &
 - k\beta\left[\frac{Z_{_{2+}}}{Z_{_{1+}}}
\cos\bigl(\varepsilon_{_{2+}}-\varepsilon_{_{1+}}\bigr)
-
\frac{Z_{_{2-}}}{Z_{_{1+}}}
\cos\bigl(\varepsilon_{_{2-}}-\varepsilon_{_{1+}}\bigr)\right], \ \ \ \ \ \ \
\label{phase_eqn_epsilon1+}\\
& & \dot \varepsilon_{_{1-}} = -k(U-c_g) \nonumber \\
& &  - k\beta\left[\frac{Z_{_{2+}}}{Z_{_{1-}}}
\cos\bigl(\varepsilon_{_{2+}}-\varepsilon_{_{1-}}\bigr)
-
\frac{Z_{_{2-}}}{Z_{_{1-}}}
\cos\bigl(\varepsilon_{_{2-}}-\varepsilon_{_{1-}}\bigr)
\right], \\
& & \dot \varepsilon_{_{2+}} = -k(-U+c_g) \nonumber \\
& &  - k\beta\left[\frac{Z_{_{1+}}}{Z_{_{2+}}}
\cos\bigl(\varepsilon_{_{2+}}-\varepsilon_{_{1+}}\bigr)
-
\frac{Z_{_{1-}}}{Z_{_{2+}}}
\cos\bigl(\varepsilon_{_{2+}}-\varepsilon_{_{1-}}\bigr)
\right], \\
& & \dot \varepsilon_{_{2-}} = -k(-U-c_g) \nonumber \\
& & - k\beta\left[\frac{Z_{_{1+}}}{Z_{_{2-}}}
\cos\bigl(\varepsilon_{_{2-}}-\varepsilon_{_{1+}}\bigr)
-
\frac{Z_{_{1-}}}{Z_{_{2-}}}
\cos\bigl(\varepsilon_{_{2-}}-\varepsilon_{_{1-}}\bigr)
\right].
\eeqa
\end{subequations}
From the analysis of the previous section,
an unstable normal mode implies a number of relationships to stand between
the quantities appearing in (\ref{eight_odes}): that (i) that $\dot\varepsilon_{_{1\pm}}=0,
\dot\varepsilon_{_{2\pm}}=0$ since the modes do not ``propagate"
but are standing in that instance, (ii)
$\varepsilon_{_{2+}}=\pi+\varepsilon_{_{1+}}$,
and
$\varepsilon_{_{1-}}=\pi+\varepsilon_{_{2-}}$,
(iii)  $Z_{_{2+}}=Z_{_{1-}}$ and
$Z_{_{1+}}=Z_{_{2-}}$.  If we assign to $\sigma$ the meaning
of a growth rate then it is related to the phase differences
according to
\[
\sigma = k\beta
\sin\Bigl(\varepsilon_{_{2+}}-\varepsilon_{_{1-}}\Bigr),
\]
in Figure 1 we have depicted $\sigma$ - in terms of
Im($c$) which is equal to $-\sigma/ik$.\par
In the extreme limit where $\beta \ll 1$,
together with the above stated
relationships,
we may establish the relative scaling behavior
between $Z_{_{1-}}$ and $Z_{_{1+}}$
and between
$Z_{_{2+}}$ and $Z_{_{2-}}$ in the unstable normal mode regime.
Taking for instance
 (\ref{phase_eqn_epsilon1+}), if one is in the
 instability range then
it immediately follows that
\beq
Z_{_{1+}} = \beta\frac{Z_{_{1-}}}{U+c_g} + \order{\beta^2},
\eeq
where we have exploited the fact that
$Z_{_{2+}}= Z_{_{1-}}$.  The scaling on
$Z_{_{2-}}$ immediately follows from
the equality of it to $Z_{_{1+}}$.  We learn from
this analysis an important clue: namely that in the limit
where $\beta \ll 1$ the main actors involved in the development of
the normal-mode instability are
the kernel waves $\zeta_{_{2+}}$ and $\zeta_{_{1-}}$ while
the other two kernel waves, $\zeta_{_{2-}}$ and $\zeta_{_{1+}}$, should
be viewed as being slaved (in a qualitative sense) to the former two.
This fact allows
us to interpret the instability with greater transparency. We devote
our attention to this
in the next section.
\subsection{The normalmode instability rationalized}
Although for this problem there are four KGW waves,
the general normal-mode analysis indicates that exponential behavior
is mainly rooted in the two individual KGWs which move against
the prevailing (background) flow.  Take for example the KGW propagating
against the flow associated with
the top interface: if one were to move into the rest frame at that level,
i.e. in a frame moving eastward with velocity $U$, then this wave would appear
to move westward. By symmetry reasons the corresponding KGW on the bottom interface would appear
to move eastward.
Thus to an observer in the laboratory frame the culprit KGWs propagate with a speed which
is slower than
the local flow.  As mentioned earlier,
we shall refer to these waves with this character as \emph{counterpropagating-KGWs}.
Taken as individual waves, i.e. free of cross-layer interactions,
the counterpopagating KGWs are those KGWs whose wavespeeds are $\pm(U-c_g)$.
Those KGWs that propagate \emph{with} the local flow, i.e.
those which have wavespeeds equal to $\pm(U+c_g)$, will be referred to
as \emph{propropagating-KGWs}.
\par
We further remind ourselves that when layers interact, the primary (eigen)modes leading to instability
come in pairs - showing either an eastward or westward propagation of the eigenmode pattern.
In general the individual eigenmodes are structures composed of all four
KGWs however we see from the arguments in the last section that when $\beta$ is small,
the eigenmodes are primarily constructed of only the two counter-propagating-KGWs
$\zeta_{_{1-}}$ and $\zeta_{_{2+}}$ while the propropagating-KGWs, namely
$\zeta_{_{1+}}$ and $\zeta_{_{2-}}$ are (in a sense) slaved to the former two.
Instability develops by which
the modes in question
show wavespeeds passing through zero.  Thus at marginality
we have two wave patterns that both appear standing to the observer in the laboratory frame.
\par
From these
observations we shall argue here that
the mechanism for instability in this problem is strongly analogous to the process
leading to the instability of CRWs as in Rayleigh's classic problem.
Thus we propose that the dynamics are governed only by the action of the
counter-propagating-KGWs for $\beta \ll 1$ (i.e. $K\gg 1$).  Since in that
case the propropagating-KGWs are slaved to the counter-propagating ones, we
posit that
the dynamical evolution of this system (for $c_{g1}=c_{g2}=c_g$)
is given by the more simpler form
\beqa
\left(\frac{\partial}{\partial t}+ik\tilde c \right)\tilde\zeta_{_{1-}} &=& -ik\beta
\tilde\zeta_{_{2+}},\label{counterKGW_top_eqn_simplified}
\\
\left(\frac{\partial}{\partial t}-ik\tilde c \right)\tilde\zeta_{_{2+}} &=& ik\beta
\tilde\zeta_{_{1-}},
\label{counterKGW_bottom_eqn_simplified}
\eeqa
where $\tilde c \equiv U - c_g$.
We note immediately that mathematically speaking,
(\ref{counterKGW_top_eqn_simplified}-\ref{counterKGW_bottom_eqn_simplified}) have the
same structure as the equations describing the evolution of interacting
kernel Rossby waves (KRWs,
cf. 9a-c of Heifetz \& Methven, 2005 \cite{heifetzCRW}).  It means to say that
these mutual interaction of KGW waves (in this limit) may be
considered and rationalized in the same manner in which interacting KRWs are
rationalized.  The familiar instability properties exhibited in the problem
of CRWs follows onto
the quality of these dynamics as well, in particular,
the co-action of hindering-helping of individual waves when instability
is present.
\par
{We may further rationalize this instability, as viewed through through
the lens of the reduced model
(\ref{counterKGW_top_eqn_simplified}-\ref{counterKGW_bottom_eqn_simplified}),
in terms of the stability criterion  of Hayashi \& Young (1987)
and Sakai (1989) \cite{hayashi_young,sakai}
referred to in the Introduction.
Inspection of the dispersion relationship
for the system (\ref{counterKGW_top_eqn_simplified}-\ref{counterKGW_bottom_eqn_simplified}),
e.g. the counter-propagating modes in Fig. \ref{total_dispersion},
shows that the those modes which become unstable propagate
opposite to each other for wavenumbers just beyond
the onset of instability.  Their Doppler-shifted frequencies
at onset (i.e. when $c=0$ for both waves) are $\pm k\tilde c$ respectively
and they are almost equal by virtue of $\tilde c$ being almost zero in the limit appropriate
for this reduced model.
And finally without the term $\beta$, which
represents the mutual interaction of these waves,
there is no chance for instability to occur.}
\section{Normal modes of a foamy layer}
\begin{figure}
\begin{center}
\leavevmode
\epsfysize=8.cm
\epsfbox{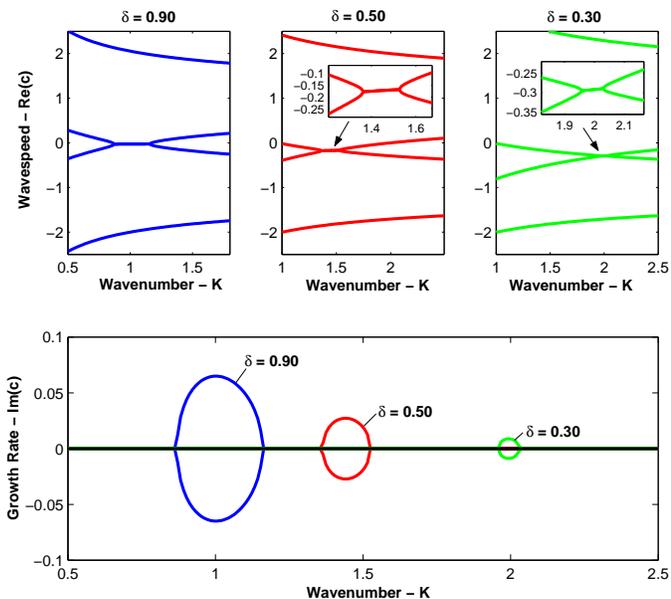}
\end{center}
\caption{{\small Wavespeeds and growth rates as
a function of wavenumber $K$ for various values
of $\delta$ at $Ri = 1$.  The top row of panels demonstrates the
wavespeeds.  Whenever instability sets in the wavespeeds
of two modes merge.  Unlike $\delta \approx 1$ theory,
the wavespeeds are non-zero and have $K$ dependence when
instability sets in.  The bottom panel shows how the growth
rate behaves as $\delta$ becomes small:
(i) the range in $K$ for which growth occurs narrows and,
(ii) and the amplitude of the maximum growth rate reduces.}}
\label{delta_plots}
\end{figure}

We consider further results assuming the two density jumps
represent the configuration of a foamy layer as discussed
in the Introduction.  In particular we will examine the
atmosphere with stratification considered in Shtemler et al. (2007).\cite{Miki_2007}
In other words we shall analyze an atmosphere in which the density of the
air $\rho_a$ is $\delta$ times the density of the foam $\rho_f$
which is, in turn, $\delta$ times the density of the
sea layer $\rho_s$.  In terms of the density formulation
given in (\ref{density_profile}) we make the following
replacements $\rho_0 \rightarrow \rho_f$ and
\[
\Delta\rho_1 \rightarrow \rho_f\frac{1-\delta}{\delta},
\qquad
\Delta\rho_2 \rightarrow -\rho_f(1-\delta),
\]
and for the velocities we have
\[
c_{g1}^2 \rightarrow \frac{c_f^2}{\delta},\qquad
c_{g2}^2 \rightarrow c_f^2,\qquad
c_f^2 = \frac{g(1-\delta)}{2k},
\]
where $0<\delta<1$.
Considerations of this system in a small vicinity
near $\delta = 1$ recovers the dynamical
behavior of the case studied in the previous section.
The dispersion relationship (\ref{dispersion_relation})
now appears as
\beqa
& & \left(\frac{c}{U}\right)^4
-\left(2 + {\cal C}_f^2\frac{1+\delta}{\delta}\right)\left(\frac{c}{U}\right)^2
-2{\cal C}_f^2\frac{1-\delta}{\delta}\left(\frac{c}{U}\right)+ \nonumber \\
& & \ \ \ \ \ \ 1-{\cal C}_f^2\frac{1+\delta}{\delta} + \frac{{\cal C}_f^4}{\delta}(1-e^{-4K}) = 0.
\eeqa
where ${\cal C}_f \equiv c_f/U$.
Inspection of the above relationship suggests that when instability
occurs, the unstable waves will have some propagation associated with them
unlike the stratification considered in the previous section.  In a similar
fashion we define the quantity $R_f$ such that ${\cal C}_f^2 \equiv R_f/K$ and where
\[
\frac{1}{F_r} = R_f \equiv \frac{gh(1-\delta)}{2 U^2},
\]
which is related to the inverse of the Froude number.  We show in
Fig \ref{delta_plots} the general trends for these circumstances
when $R_f=1$.
A non-zero value of $\delta$ means that when instability sets
in, it is no longer stationary as in the previous section, but
the modes in question are now propagatory.
In the range of unstable wavenumbers the propagation speed is
is also a function of $K$.  For $R_f$ held fixed, as $\delta$ approaches zero the
range of unstable wavenumbers gets smaller, the location of the
unstable band shifts to larger wavenumbers and the peak growth
rate in this unstable band similarly get smaller.\par
\begin{figure}
\begin{center}
\leavevmode \epsfysize=8.cm \epsfbox{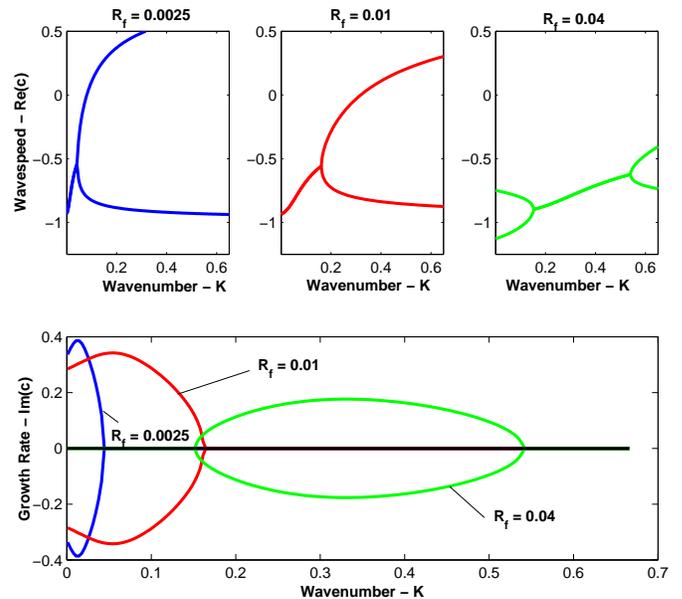}
\end{center}
\caption{{\small Wavespeeds and growth rates for a foamy
layer as a function of
wavenumber $K$ for various values of $R_f$ at $\delta = 0.03$.  The
top row of panels demonstrates the wavespeeds of those modes
demonstrating growth and decay only.  The peak
growth rates get larger for smaller values of the bulk
Richardson number.}} \label{Rf_plots}
\end{figure}

In Fig. \ref{Rf_plots} we display the results of fixing $\delta$
and varying the bulk Richardson number $R_f$.  This time, however,
we use numbers that figure to be relevant to the problem
of hurricanes.  Because the density ratio between water and air is
one thousand to one we assume that $\delta = \sqrt{0.001} \approx 0.03$.
Following Shtemler et al. (2007)\cite{Miki_2007} we take (as a rough order
of magnitude estimate) for the foam layer thickness the figure $h=1$m.
Because, empirically speaking, strong winds over water follow a logarithmic
profile Powell et al. (2003)
\cite{powell} generate fits of this form to the data accumulated during hurricane
over-flight missions.  For hurricane force winds (i.e. winds
exceeding 34 m$\cdot$s$^{-1}$ at 1km above the sea surface)
the wind speed at 1m above the sea surface
is approximately 20 m$\cdot$s$^{-1}$ as inferred from Figure 2
of Powell et al. (2003)\cite{powell}. Thus we assume this
figure for $U$.  Roughly
speaking, it means that
the bulk Richardson number for these conditions is $R_f \sim 0.01$.
The results depicted in Fig. \ref{Rf_plots} are for three
values of $R_f = 0.04,0.01,0.0024$.  The trends appearing
are that the peak growth rate increases as $R_f$ is made smaller although
the unstable range in $K$ begins to narrow as well.  Just as in
the results of the previous section (e.g. see Fig. \ref{linear_instability_diagram})
there exists a critical value of $R_f(\delta)$ below which $K=0$ is also unstable.
In general this is probably a pathology of the theory and one
ought not to take too seriously the results at $K=0$.  However, once
$R_f$ dips below this value the range of unstable wavenumbers
narrows.  We note finally that the first wavenumber to be unstable
at $R_f = 0.01$ occurs at $K \approx 0.15$.  We shall make a remark
about this and its relationship to hurricane data in the
next section.

\section{Summary and Discussion}
\subsection{Recapitulation}
We have studied how KGWs interact across two
layers in a medium with globally constant shear.
KGWs, along with KRWs \cite{heifetzCRW}, are
dynamical phenomena forming a subclass
of KRGWs developed in Harnik et al. (2007)\cite{HHUL07}.
In terms of the formalism developed, we have found that
the interaction of KGWs under these conditions bear
significant similarity to KRWs especially when instability
sets in.  For KGWs, the presence of a jump
in the background density at some
layer means
that these places support the creation of vorticity
which ultimately relates to the interpretation that
such configurations are sources of baroclinic torques
\cite{UH_07}.  Unlike
KRWs, where a jump in the background shear produces
a single propagating Rossby edge-wave,
an isolated jump in density creates two surface
gravity waves propagating with equal and opposite
directions.  These cohabitating (in the sense
that they exist on the same layer) counter-propagating
gravity waves can be cast into a
a set of new variables which explicitly shows that
the wave pair do not interact.
\par
The KGW kernel is
composed of the measure of the edge-wave vorticity and
the amount which the surface has been displaced
from equilibrium.  Viewed in terms of this combination
of physical variables, we
have demonstrated how a single wave mechanically
operates as it propagates.  The vorticity field
corresponds to a velocity field around the disturbed
surface while the disturbed surface indicates how
further vorticity is to be generated ahead of the
peak of the wave.  The disturbed surface creates
this vorticity (in the sense of the mechanical
description we present) because it looks like a
displaced lever arm - when let free such a lever
arm has the tendency to rotate about its equilibrium
point and generate vorticity as a consequence.  The
true wave executes these steps continuously in a fluid
way.
\par
The analysis of the interaction of KGWs on two layers
was facilitated here by assuming the density jumps
to be the same across each layer which
results in a normal mode problem that is analytically
tractable.  Because of the shear profile, the top
layer moves with speed $U$ while the bottom
layer moves with speed $-U$.
In this case one has four normal modes
(a co-habitating pair on each layer).  Normal mode
instability arises primarily through the interaction
of two modes from each layer - these modes
are ones which, in the absence of mutual interaction,
would propagate against the background flow of the layer.
The remaining two modes contribute less and less to the
development of the instability when the layer separation
(or, equivalently, the horizontal wavenumber) gets large.
In this large separation limit, the resulting effective
equations have the same
mathematical form as the ones describing
the (exact) development of unstable CRWs \cite{heifetz99}.
This means to say that the mechanism behind this instability
is qualitatively similar to the way
instability emerges in the problem of CRWs.
\par
\subsection{As being a CRW analog}
In a sense
the instability is the analog of the CRW instability and
it is for this reason why we refer to this mode interaction
process as the instability of counter-propagating gravity waves.
Indeed the KGW may be interpreted as a (delta-function) vortex structure
which propagates at some speed on the layer in question.
Whereas in KRWs the source of the vorticity is the jump
in the background vorticity, in KGWs the source is the
baroclinicity that arises when the jump layer is perturbed.
The ensuing dynamics are qualitatively similar in all other
major respects.  It is also true that both left and right going
waves become excited on a layer when a single KGW propagates along
the other.  In this way the conception and rationalization of
the dynamics of four KGWs is more complicated.  But, and
we reiterate this point, the basic qualitative features of
the instability involves the dynamical influence between only
two of the KGWs, {namely those that counterpropagate with respect
to the background flow at their respective interfaces},
and the processes at work resemble
the mechanism responsible for the instability of CRWs.

\subsection{The normal-mode instability, its rationalization
and relationship to the Miles-Howard Theorem}
 {As the disturbances considered here preclude the possibility
of an inflexion point instability (i.e. the Rayleigh criterion), the character
of disturbances falls under the purview of the Miles-Howard
theorem.
The Miles-Howard theorem \cite{miles61,howard61} for the stability of
plane-parallel stratified shear flows states that a sufficient
criterion for stability of normal-mode disturbances
is that the Richardson number (J) of the flow be
everywhere greater than $\sfrac{1}{4}$ in the
domain.  That is to say that a necessary condition
for instability is that somewhere in the flow
\beq
{\rm J} \equiv -\frac{g}{\rho}{\frac{d\rho}{dz}}\cdot \frac{1}{(dU/dz)^2} < \frac{1}{4},
\label{miles-howard}
\eeq
where the quantities defining J are in dimensional units.
Because this is a plane-Couette flow $dU/dz$ is everywhere defined and not equal to
zero.    For the ``stably"
stratified fluid we have considered here, i.e. $-(g/\rho)d\rho/dz > 0$,
J would then be described by two delta-functions, with positive coefficients,
centered on the locations of the density discontinuity.
This means to say that J is zero everywhere except at the points
where the density discontinuity occurs.
At the very least, the instability uncovered here satisfies
the necessary Miles-Howard criterion for normal-mode instability
as J is less than $\sfrac{1}{4}$ in substantial parts
of the flow.}
\par

\par
 {What is interesting to reflect upon is that this instability might
be the simplest kind possible in a shear flow \emph{absent
vorticity waves}. Regular (inviscid) plane-Couette flow, though satisfying
the necessary condition for instability, is however stable as
is a plane-Couette flow with a single density interface.
All that is needed for a plane-Couette profile to exhibit
normal-mode instability
is for there to exist a second density interface.}
\par
 {Given the previous discussion it seems to us that
normal-mode instability in shear flows
merely requires the presence of counter-propagating waves
of any sort. Indeed a cursory review of some previous work reveals: (i)
the classic instability emerging from
the model problem investigated by Rayleigh\cite{Rayleigh1880}
has been reinterpreted as arising out of the interaction of phase-locked
counter-propagating
vorticity waves \cite{bretherton,heifetz99} (i.e. Rossby edge waves), (ii) a simplified restricted
version of Holmboe's instability
\cite{holmboe62} has been shown
to be emerging out of the interaction of a pair of counter-propagating waves
in which one is a vorticity wave and the other is a gravity wave\cite{BM94} and,
(iii) here we have demonstrated yet another instability
borne out of the interaction of counter-propagating
waves which are comprised, in this instance, of
phase-locked gravity waves.}
\par
This leads to the apparently counterintuitive conclusion that
a convectively stable stratification can destabilize
a flow profile that is stable in the unstratified case.
It is nevertheless important to recall here that this result is not in
contradiction with conventional linear stability criteria.
Indeed, in the presence of stratification, one of these criteria is the
Miles-Howard criterion found in (\ref{miles-howard}) which places no limitations
on the wind curvature. It means that there is a much larger array
of wind profiles that can be unstable in the stratified case
than in the unstratified one.
\par

\subsection{Foamy layers and a conjecture}
\par
We observe that the range of unstable wavenumbers is finite in the
case where the density jumps are the same (Section IV). But, in
comparison, both the wavenumber range and the maximum growth rate
of the instability when considered in a foamy layer (Section V)
reduces as the the density parameter describing the foam $\delta$
gets small.  However consideration of parameters having more
relevance to real atmospheric phenomenon reveals some interesting
predictions.  When numbers appropriate for observed hurricanes are
utilyzed\cite{powell,Miki_2007} we find that they correspond to
bulk Richardson numbers ($R_f$) in the vicinity of $0.01$.
Inspection of the growth rates show that the wavenumber for the
onset of instability is $kh \approx 0.15$ in which $h$ is the
nominal height of the foam layer. Assuming, as we have, that
$h=1$m it means that the wavelength of this unstable mode is
approximately $\lambda \sim 45$m.  An inspection of Figure 4a in
Powell et al. (2003)\cite{powell} shows a photograph of the sea
from a quarter kilometer height during a hurricane whose
windspeeds, as measured at the height of the aircraft, are
approximately $45$ m$\cdot$s$^{-1}$. The image shows the foamy
patches and the length scales of the groupings that appear is
approximately $30$m in length. Figure 4b of the same work shows
the foamy patches after the windspeeds have increased to $55$
m$\cdot$s$^{-1}$ and one sees clearly that the foam patches have
covered the entirety of the sea surface.  Since larger values of
the windspeed correspond to smaller values of critical
wavenumbers, the theory developed here shows at least the same
trend in the observed photographs.\par
Of course there are obvious
shortcomings of applying the theory developed here to the
circumstances noted for this hurricane observation: (i), that the
height of the foam layer was assumed since there are no direct
observations to that end from the data available, (ii) it is not
at all clear if there exists a ``laminar" foam layer from which a
secondary instability develops creating the frothy foam layer
observed in these photographs and, (iii) the assumption of a
linear velocity profile in place of a logarithmic one is clearly a
gross simplification.  The latter assumption is probably
satisfactory on the scale of the foam layer but certainly breaks
down if one tries to extend it much beyond a logarithmic scale
height of the layer thickness (e.g. $10$m or higher if $h=1$m).
Thus although the apparent consistency observed between this
simple theory and the observations made is not proof that the
theory developed here is relevant to such complicated processes
(i.e. hurricane-sea interfaces), it is at least encouraging that
this theory makes predictions which are of at least the same order
of magnitude as the structures observed.  As such we treat this
discussion as being purely conjecture at this stage.
\begin{acknowledgments}
This research was supported by BSF grant 2004087
and ISF grant 1084/06.
Nili Harnik was also supported by the European
Union Marie Curie International reintegration Grant
MIRG-CT-2005-016835.
Part of the work was done when EH was a Visiting
Professor at the Laboratoire de M\'et\'eorologie Dynamique at the
Ecole Normale Sup\'erieure.
\end{acknowledgments}

\appendix
\section{Derivation}
We show here the step-by-step derivation that ultimately leads to the
equation set (\ref{hat_q1_eqn}-\ref{hat_zeta2_eqn}).  The basic density
profile and its derivatives, written in terms of $b$, is itemized
here,
\beqa
& & \frac{d\bar b}{dz} = \frac{g\Delta\bar\rho_1}{\rho_0} \delta(z-z_1) +
\frac{g\Delta\bar\rho_2}{\rho_0} \delta(z-z_2), \\
& & \frac{\partial b}{\partial x} = -\frac{g\Delta\bar\rho_1}{\rho_0}
\frac{\partial \zeta_1}{\partial x}\delta(z-z_1)
-\frac{g\Delta\bar\rho_2}{\rho_0}\frac{\partial \zeta_2}{\partial x} \delta(z-z_2), \ \ \ \ \
, \\
& & \left(\frac{\partial}{\partial t} + \bar U\partial_x\right) b =
-\frac{g\Delta\bar\rho_1}{\rho_0} \delta(z-z_1) \left(\frac{\partial}{\partial t} + \bar U\partial_x\right) \zeta_1
\nonumber \\
& & \ \ \  -\frac{g\Delta\bar\rho_2}{\rho_0} \delta(z-z_2) \left(\frac{\partial}{\partial t} + \bar U\partial_x\right) \zeta_2
.
\eeqa
these expressions are inserted into the governing equations
(\ref{q_eqn}-\ref{zeta_eqn}),
\beqa
& & \left(\frac{\partial}{\partial t} + \bar U\partial_x\right)q =
(g/\rho_0)\frac{\partial \zeta_1}{\partial x} \delta(z-z_1) \nonumber \\
& & \ \ \ \ \ \ \ \ \ \ \ \ \  +(g/\rho_0)\frac{\partial \zeta_2}{\partial x} \delta(z-z_2)
, \label{vorty_2jump}
\\
& & \left(\frac{\partial}{\partial t} + \bar U\partial_x\right)
\left[\frac{g\Delta\rho_1}{\rho_0} \zeta_1 \delta(z-z_1) +
\frac{g\Delta\rho_2}{\rho_0} \zeta_2 \delta(z-z_2)\right] \nonumber \\
& & \ \ \ \ \ \ \ \
=  w\frac{g\Delta\rho_1}{\rho_0}  \delta(z-z_1) + w\frac{g\Delta\rho_2}{\rho_0}  \delta(z-z_2)
\label{continuity_2jump}
.
\eeqa
From inspection above it is clear that the vorticity of
the system takes on delta function character at all of
the jumps of the system.  This then means applying the ansatz
\[
q = \hat q_1(x,t)\delta(z-z_1) + \hat q_2(x,t)\delta(z-z_2).
\]
Collecting all terms of like delta functions,
 one may set each of the coefficients
to zero
because of the mutual ``orthogonality" of the delta functions.
In a sense this system has been
discretized.  This results in the following,
\beqa
\left(\frac{\partial}{\partial t} +  U\frac{\partial}{\partial x}\right)\hat q_1 &=&
h N^2_{_1}\frac{\partial\zeta_1}{\partial x}, \label{q1_eqn}\\
\left(\frac{\partial}{\partial t} +  U\frac{\partial}{\partial x}\right)\zeta_1 &=&  w(h,x), \\
\left(\frac{\partial}{\partial t} -  U\frac{\partial}{\partial x}\right)\hat q_2 &=&
h N^2_{_2}\frac{\partial\zeta_2}{\partial x}, \\
\left(\frac{\partial}{\partial t} -  U\frac{\partial}{\partial x}\right)\zeta_2 &=&  w(-h,x).
\label{zeta2_eqn}
\eeqa
To recover (\ref{hat_q1_eqn}-\ref{hat_zeta2_eqn}) we apply the Fourier ansatz (\ref{fourier_ansatz})
to (\ref{q1_eqn}-\ref{zeta2_eqn}).
With which it immediately follows that the streamfunction field $\psi$ (subject to the
usual jump conditions) is the familiar form,
\beq
\psi = -\left(\frac{\tilde q_1}{2k}e^{-k|z-h|}
+\frac{\tilde q_2}{2k}e^{-k|z+h|}\right)e^{ikx} + {\rm c.c.}.
\label{hat_psi_solution}
\eeq
\par

\end{document}